# Topological-charge dependence of radiation torque for an acoustic-vortex spanner

Yuzhi Li[1], Gepu Guo[1], Juan Tu[2], Qingyu Ma[1, 2*], Xiasheng Guo[2], Dong Zhang[2†], Oleg A. Sapozhnikov[3,4], and Thomas J. Matula[3]

[1] School of Physics and Technology, Nanjing Normal University, Nanjing 210023, China

[2] Institute of Acoustics, Nanjing University, Nanjing 210093, China

[3] Center for Industrial and Medical Ultrasound, University of Washington, Seattle, WA 98105, USA

[4] Department of Acoustics, Physics Faculty, Moscow State University, Moscow 119991, Russia

Corresponding authors

Qingyu Ma : *maqingyu@njnu.edu.cn

Dong Zhang : †dzhang@nju.edu.cn

School of Physics and Technology

Nanjing Normal University

1 Wenyuan Road, Xianlin District

Nanjing 210023, China

Phone: 86-25-85891905

Mobile: 86-13770847142

# Abstract

Based on the analyses of the wave front and the wave vector of an acoustic-vortex (AV) spanner generated by a circular array of point source, the principle of object rotation is investigated through the calculation of the orbital angular momentum (OAM) and the radiation torque for AVs with various topological charges. It is demonstrated theoretically that the rotation of the axisymmetric disk centered on an AV spanner is mainly driven by the exerted radiation torque without the contribution of the OAM transfer. The radiation torque on a small-radius object is inversely associated with the topological charge in the center of the AV spanner, and it is enhanced significantly for a larger AV with a higher topological charge. The special case of the radiation torque proportional to the topological charge might be realized when the disk radius is much larger than the wavelength, in which case the acoustic power of the AV spanner can be absorbed as much as possible. With the established sixteen-source experimental setup, the radial pressure distributions of AVs with different topological charges measured at the frequency of 1.3 kHz in air agree well with the simulations. The topological-charge dependence of the radiation torque for AVs is also verified by the quantitative laser-displacement (angle) measurements for disks with different radii. The favorable results demonstrate that, for the object covering the vortex center of an AV spanner, the total OAM transfer might be 0 and the object rotation is contributed by the acting moments. Thus, the radiation torque of an AV spanner is more applicable than the OAM for describing the driving capability of object rotation, and it can be used as an effective tool in clinical applications to noninvasively manipulate objects with a feature size at the wavelength-scale (e.g. kidney stone in lithotripsy) inside body.

**Keywords**

Acoustic vortex, radiation torque, topological charge, orbital angular momentum, object rotation

PACS : 43.25.Qp, 43.60.Fg, 43.38.Hz

# I.  INTRODUCTION

Based on the concept of optical vortices [1-8], the acoustic vortex (AV) was proposed to be generated by helical wave fronts with indeterminate phases and pressure nulls along the propagation axis [9-17], possessing the capability to capture and manipulate particles with the radiation force and the orbital angular momentum (OAM). The circular phase variation around the vortex core in the transverse plane can be described by $2\pi l$ for the topological charge *l* [16, 18], indicating how many phase twists the acoustic wave undergoes in one circle. Small objects can be driven to rotate around the vortex center and accumulate at the center axis along the acoustic beam, which has been applied in the fields of object alignment and particle manipulation [19, 20]. Compared with optical vortices, AV beams can be generated in deep tissues, thus exhibiting the potential for broad applications in minimally invasive treatments and targeted drug delivery [21] in biomedical area.

In the past decades, the application of AVs has attracted more and more attentions in the discussion of acoustic tweezers, acoustic spanner and acoustic trapping. Based on the solution of the Helmholtz equation, the intrinsic OAM properties of AV beams were explained by Lekner [22] and the corresponding OAM was calculated *via* quantum mechanics. Zhang [23] pointed out that, for a monochromatic nonparaxial acoustic vortex beam, the acoustic radiation torque exerted on an axisymmetric object centered on AV axis could be calculated as the integral of the time-averaged flux of OAM density over the total field, which was proportional to the power absorbed by the object with a factor of $l/\omega$ for the angular frequency $\omega$. An interpretation of the AV in free space was proposed by Volke-Sepúlveda [24], and the difference was evaluated for AVs with varied topological charges (*viz.*, *l*=1 and 2). The generation of AV was experimentally demonstrated by the steady deflection of a hanging disk using a four-source array at a fixed phase difference of $\pi/2$. A torsion pendulum was used to study the OAM transferred to hanging disks of several sizes for AVs with $l$=1 and 2. In Skeldon's study [25], the acoustic wave with helical phase fronts generated by a circular array of loudspeakers was predicted to have an OAM with the energy ratio of $l/\omega$. The OAM of an AV spanner at the frequency of 500 Hz was confirmed by measuring its transfer to a suspended acoustic absorbing tile (diameter =60 cm, less than the wavelength of 68 cm) only for $l=\pm1$, and the helical phase structure of the acoustic beam was also confirmed by measuring the Doppler shift of a rotating observer. In addition, a quantitative test of the OAM transferred to an acoustic absorbing object immersed in a viscous liquid was performed by Anhauser [26]. A steady spinning frequency was used to analyze the balance of the radiation torque exerted by the AV beam and the viscous torque exerted by the host fluid. A phase-coded approach was proposed by Yang *et al.* [27] to generate an AV beam with controllable topological charge and spiral direction of phase twists. The maximal topological charge of the AV beam generated by a circular array of $N$ point sources was proved to be Fix[($N$-1)/2], where Fix($x$) rounded the element $x$ towards zero.

In the previous studies done by Lekner and Volke-Sepúlveda [22, 24], the capability of object manipulation was reported to be determined by the OAM transferred from the AV, which was proportional to the topological charge as $\boldsymbol{L}_z = \rho(\boldsymbol{r} \times \boldsymbol{v}_\perp) = \pm(\rho l \varepsilon)/\omega$, where $\boldsymbol{v}_\perp$ was the particle velocity in the transverse plane, $\rho$ was the density of the medium, $\omega$ was the angular frequency of the wave, and $\varepsilon$ was the energy density of the acoustic field. It was clear that, for a fixed object, faster object rotation could be produced by the greater OAM transferred from the AV with a higher topological charge. However, Volke-Sepúlveda observed that the radiation torque of the AV for *l*=2 was smaller than that for *l*=1 when the radius of the suspension disk was smaller than 4.0 cm[24], which is contrary to the factor of $l/\omega$, without any theoretical analysis provided to explain the phenomenon. Therefore, further investigation on the topological-charge dependence of the OAM or the radiation torque for an AV spanner plays a key role in object rotation.

Also as we know, the capability of object manipulation of an AV spanner is determined by the rotational characteristics of the wave vector $\boldsymbol{k}$. In an ideal medium without viscosity and dissipation, the particle velocity $\boldsymbol{v}$ of the acoustic wave radiated from a point source is along the propagation direction, which can be calculated by $\boldsymbol{v} = -\nabla p/(i\omega\rho)$. Although $\boldsymbol{v}$ is irrotational, the acoustic radiation force $\boldsymbol{F}$ exerted on the object is the time-average of acoustic excess pressure [28-30] with the direction pointing to the normal direction of the wave front. Similar to the wave vector of the AV beam, the acoustic radiation force is rotational for $\nabla\times\boldsymbol{F}\neq0$. As a result, the acoustic radiation force and the corresponding radiation torque of an AV spanner can be applied properly to investigate the property of object rotation instead of the OAM transfer.

In this paper, to overcome the limits of the irrotationality and the periodicity of the particle velocity, the formulae of the acoustic radiation force and the radiation torque of an AV spanner generated by a circular array of point sources were derived based on the analyses of the wave front and the wave vector of the helical AV beam. For the disk centered on an AV spanner, the total OAM transfer was proved to be zero, demonstrating that the disk rotation was the result of the acting moments. A sixteen-source AV generation system and a laser-displacement (angle) indicator device were established to measure the radiation torque quantitatively. The consistent results of numerical simulations and experimental measurements of the acoustic pressure and the radiation torque for AVs showed that the radiation torque exerted on a small-radius object was inversely associated with the topological charge in the center of an AV spanner, and it is enhanced significantly for a larger AV spanner with a higher topological charge. The special case of the radiation torque proportional to the topological charge [23] might be realized when the disk radius is much larger than the wavelength, in which case the acoustic power of the AV spanner can be absorbed as much as possible. For an object covering the vortex center, the total OAM transfer might be 0 and the object rotation is mainly produced by the exerted radiation torque of the AV spanner. The favorable results suggested that the radiation torque of an AV spanner is more applicable than the OAM transfer in describing the driving capability of object rotation, and can be used in biomedical areas as an effective method to manipulate the objects with a wavelength-scale featured size (e.g. kidney stone in lithotripsy).

## II. PRINCIPLE AND METHOD

The phase-coded approach [27] for a circularly distributed source array is employed according to the schematic diagram as shown in Fig. 1. $N$ acoustic point sources are positioned on a circumference (radius $a$) with a spatial angle difference $\Delta\varphi = 2\pi/N$. In the cylindrical coordinates, the position of the $n^{\text{th}}$ source can be described as $(a,\varphi_n,0)$ with $\varphi_n = 2\pi(n-1)/N$. When the sources are excited by $N$ sinusoidal waveforms with a phase difference $\Delta\phi = l\Delta\varphi$, the acoustic pressure at $(r,\varphi,z)$ produced by the $n^{\text{th}}$ source can be

written as

$$p_n(r,\varphi,z,t) = (A_0/R_n)\exp(i\omega t)\exp(-ikR_n)\exp(\pm i\phi_n), \tag{1}$$

where $A_0$ is the acoustic pressure of the sources, $k = \omega/c$ is the wave number for the acoustic speed $c$, $R_n$ is the propagation distance from $(a,\varphi_n,0)$ to $(r,\varphi,z)$, and $t$ is the transmission time. Thus, the total acoustic pressure at $(r,\varphi,z)$ can be calculated by

$$p(r,\varphi,z,t) = \sum_{n=1}^{N}(A_0/R_n)\exp(i\omega t)\exp(-ikR_n)\exp(\pm i\phi_n). \tag{2}$$

It is clear that $p(r,\varphi,z,t)$ comprises the information of pressure amplitude and phase, which can be used to describe the helical structure of the wave front by $\Phi(r,\varphi,z,t) = \arctan[\mathrm{Im}(p)/\mathrm{Re}(p)]$. Then, through gradient calculation for the phase distribution $\nabla\Phi = (\partial\Phi/\partial x, \partial\Phi/\partial y, \partial\Phi/\partial z)$, the normal direction of the wave front at $(r,\varphi,z)$ can be achieved to describe the direction of the acoustic radiation force.

In an acoustic field, the acoustic radiation force incident on a point can be expressed by $\mathrm{d}\boldsymbol{F} = \left\langle \rho\boldsymbol{v}(\boldsymbol{v}\cdot\boldsymbol{n}) + \left(\frac{p^2}{2\rho c^2} - \frac{\rho\boldsymbol{v}\cdot\boldsymbol{v}}{2}\right)\boldsymbol{n} \right\rangle \mathrm{d}S$ [28, 29], where $\langle\cdot\rangle$ is the time-averaged calculation over the period $2\pi/\omega$, and $\boldsymbol{n}$ is the unit vector of the acoustic radiation force, which is perpendicular to the surface element $\mathrm{d}S$. Lee [28] concluded that the first term $\rho\boldsymbol{v}(\boldsymbol{v}\cdot\boldsymbol{n})$, known as the Reynold's stress, was originated from the momentum equation and described the nonlinear convection current in the Eulerian system, representing the time-averaged transport of the momentum density with the velocity. Mitri [31] also reported that the Reynold's stress was produced by changing from the Eulerian to the Lagrangian reference frame due to the motion of the fluid particle at the boundary, and the nonlinear term was negligible in small amplitude acoustic vibrations. Therefore, for the object in an AV spanner, the radiation torque can be calculated by the integral of all surface elements as $M = \iint_S r \times dF$, where $dF = \left( \frac{\langle p^2 \rangle}{2\rho c^2} - \frac{\rho \langle v \cdot v \rangle}{2} \right) n dS$ represents the acoustic radiation force exerted on the surface element $dS$. Since the component along the $z$ direction of the acoustic radiation force does not contribute to disk rotation, $n$ can be replaced by its projection onto the transverse plane as $n_\perp = \left( \frac{\partial \Phi}{\partial x}, \frac{\partial \Phi}{\partial y} \right) \bigg/ \left\| \left( \frac{\partial \Phi}{\partial x}, \frac{\partial \Phi}{\partial y} \right) \right\|$. Hence, the radiation torque exerted on the entire object can be calculated by

$$M = \iint_S r \times \left( \frac{\langle p^2 \rangle}{2\rho c^2} - \frac{\rho \langle v \cdot v \rangle}{2} \right) n_\perp dS. \tag{3}$$

It can be seen that, besides the acoustic pressure, the acoustic radiation torque of the AV spanner is also influenced by the angle between the normal direction of the wave front and the observation plane. For an AV spanner with the fixed parameters of array radius, source number and angular frequency, the exerted radiation torque is determined by the distribution of the acoustic pressure and the structure of the helical wave front, which are also influenced by the topological charge.

## III.   SIMULATION AND EXPERIMENT

Numerical studies were conducted for $N$=16, $a$=30 cm and $f$=1.3 kHz in air with $c$=342 m/s and $\Delta\varphi = 2\pi l / 16$. Considering the theory proposed by Yang [27], the distributions of AVs for $l$=1 to 7 were calculated. The axial pressure profiles (line 1) and the cross-sectional distributions of pressure (line 2) and phase (line 3) for AVs with $l$=1 to 4 (column 1 to 4) at $z$=15 cm are presented in Fig. 2. Circular pressure distributions with obvious phase spirals and center pressure nulls demonstrate the formation of AVs. As shown in Fig. 2(1-$l$), for each $l$, the radius of the AV at the source plane is the smallest, and an expanded AV can be generated at a longer transmission distance with a decreased peak pressure ($P_P$). At a fixed distance, a bigger radius of the AV with more divergent energy is produced for a higher topological charge. With the increase of the topological charge, the radii of the peak pressure ($R_{PP}$) and the valley pressure ($P_V$, $R_{PV}$) increase accordingly with an expanded vortex center (low-level pressure region). Corresponding to Fig. 2(2-*l*), an anti-clockwise phase spiral around the vortex core is displayed clearly in Fig. 2(3-*l*) with a phase variation of $2\pi l$. Furthermore, the phase variation in the center AV (*viz.*, $r<R_{PP}$) is mainly along the tangential direction, which is greater than that of the surrounding AV (*viz.*, $r>R_{PV}$). Compared to the center AV, the contribution of the surrounding AV on object rotation is relatively lower due to the weaker tangential component of the acoustic radiation force. Therefore, $R_{PV}$ can be considered as the effective radius of an AV spanner for rotation manipulation, which is clearly indicated by the phase circles in Figs. 2(3-1) and 2(3-2) for *l*=1 and 2. With the radial pressure distributions of AVs as shown in Fig. 3 for *l*=1 to 7, one can find that, with the increase of the radius, the acoustic pressure increases from the vortex core (*r*=0) to $P_P$ at $R_{PP}$, and then it goes down to $P_V$ at $R_{PV}$. For *l* =1 to 7, $R_{PP}$ increases monotonically from 8.4 to 34.0 cm with a decreased $P_P$ from 1 to 0.46. Due to the influence of the acoustic sources, only two $P_V$s at $R_{PV}$=17.4 and 24.0 cm are formed for *l*=1 and 2. Whereas, for *l*>3, the acoustic pressure drops monotonically from its $P_P$ without the generation of $P_V$ in the observation range.

    A disk centered on the vortex center was assumed to be hung horizontally above the source plane at *z*=15 cm to measure the radiation torque of the AV spanner. By applying Eq. 3, the dependences of radiation torque on disk radius for AVs with *l*=1 to 7 were calculated as

plotted in Fig. 4. With the increase of the disk radius, the radiation torque of the AV spanner increases accordingly with the growth rate determined by the acoustic pressure and the direction of the wave vector, as well as the size of the disk. In the center region of a small AV spanner with a little topological charge, the radiation torque increases rapidly due to the higher acoustic pressure. For the existence of $P_V$, a growth rate of almost zero can be formed after $R_{PV}$. While for the AV spanner with a larger topological charge, the radiation torque in the center region is much lower, and it increases continuously with an enhanced growth rate due to the absence of $P_V$ in the observation range. Thus, various radiation torque sequences are produced for different disk radii as shown in Fig. 4. In the center region of AVs ($r$<15.3 cm), a radiation torque sequence $S_M$(1, 2, 3, 4, 5, 6, 7) is formed, indicating that the radiation torque of the AV spanner is inversely associated with the topological charge. Due to the decline of the acoustic pressure after $P_P$, the growth rates of radiation torque for AVs with $l$=1 and 2 decrease to zero gradually after the corresponding $P_V$. Meanwhile, rapid growths of radiation torque for AVs with $l$>2 are observed along with the increase of the disk radius because of the continuous increase of the acoustic pressure. When $r$=15.3, 18.5, 22.3, 23.6 and 27.3 cm, the radiation torque sequences are changed to $S_M$(2, 1, 3, 4, 5, 6, 7), $S_M$(2, 3, 1, 4, 5, 6, 7), $S_M$(2, 3, 4, 1, 5, 6, 7), $S_M$(3, 2, 4, 1, 5, 6, 7) and $S_M$(3, 4, 2, 5, 1, 6, 7), respectively. In addition, the special sequence $S_M$(7, 6, 5, 4, 3, 2, 1) of the radiation torque proportional to the topological charge [22-24] does not appear in the observation range, and it might be realized when the disk radius is much larger than the wavelength to absorb as much acoustic power of the AV spanner as possible, which is consistent with the conclusion claimed by Zhang and Marston [23]. Therefore, for a small disk with the radius less than the half wavelength, a larger radiation torque can be generated in the center of an AV spanner with a smaller topological charge. With the increase of disk radius, an enhanced radiation torque can be produced by the expanded AV spanner with a larger topological charge, thus exhibits a stronger rotation effect.

In order to verify the formation of AVs and measure the extent of the object rotation, an experimental system, as illustrated in Fig. 1, was established with the parameters consistent with those used in the simulations. Sixteen speakers (diameter 8 cm, power 10 W) with cylindrical waveguides (radius 1 cm) were distributed on a circumference ($a$=30 cm) to generate sixteen point sources. Sixteen sinusoidal signals with controllable phases [27] were sent out by the wave generators, and then the power was amplified to drive the speakers to generate AVs above the source plane. The topological charge of the AV beam was adjusted by the phase difference $l\pi/8$ for $l$= -7 to 7. Controlled by the motion controller (Newport ESP301, Newport Corporation, USA), the radial pressure distributions were measured by the stepper motor (Newport M-ILS250, Newport Corporation, USA) with a mini-microphone (diameter 4 mm, Panasonic WM-61B102C, Japan), which were collected by the digital oscilloscope (Agilent DSO9064A, Agilent Corporation, USA).

In order to measure the radiation torque of the AV spanner accurately, a laser-displacement (angle) measurement system was developed as shown in Fig. 5. Under a coaxially hanging circular acrylic disk, a sponge layer with 2-cm thickness was adhered to absorb acoustic waves from below to reduce acoustic reflection. A soft cotton thread was connected to the center of the disk to provide weight support, and an elastic rubber hose with a fixed length was applied to measure the radiation torque exerted on the disk. In addition, a laser beam emitted by a laser diode fixed at the center of the disk was used to indicate its deflection angle by the displacement, which was measured at 5-m away from the center of the disk. Driven by the AV spanner, an opposite resistance torque could be generated by the twisted rubber hose. When it reached and maintained torque balance, the radiation torque exerted on the disk could be evaluated by the approximate linear relation $M = -\kappa\theta$, where $\kappa$ was the rotational stiffness of the rubber used and $\theta$ was the deflection angle.

Due to the range limit of the stepper motor, several radial pressure distributions were measured from 3 to 25 cm at the step of 0.3 mm for AVs with $l$=1 to 4, which are plotted in Fig. 2. It is clear that the experimental results are in good agreements with the corresponding simulations, especially for the similar locations and amplitudes of $P_P$ and $P_V$. For each $l$, the acoustic pressure in the center region of the AV spanner is relatively lower and it increases accordingly with the increase of the radius until reaching its $P_P$. In addition, for the generated AVs with $l$=1 to 4, the deflection angles were measured for four disks with the radii of 13, 18, 20 and 26 cm. The experimental results of the measured radiation torque were normalized by the maximum displacement of 37.5 cm (deflection angle 4.29°) for $r$ = 26 cm and $l$ = 3, which are plotted in Fig. 4. It is obvious that the measured deflection angles agree well with the simulation curves. The sequences of radiation torque for the four disks are $S_M$(1, 2, 3, 4), $S_M$(2, 1, 3, 4), $S_M$(2, 3, 1, 4) and $S_M$(3, 2, 4, 1), respectively. Therefore, the consistent results demonstrate that the disk deflection can be applied to evaluate the radiation torque of an AV spanner accurately.

## IV. DISCUSSION

Because of the emphasis on optical vortices indicated in previous studies, the OAM transfer was often used to describe the characteristics of AVs. However, in practical applications, the object to be manipulated cannot be regarded as a theoretical mass point when its size is not far less than the wavelength, and it is meaningless to discuss the OAM only at one point. Through integral calculation for the OAM, there must be an area with a total OAM to be 0. Consequently, it is unable to provide a reasonable explanation for the object rotation by only considering the OAM transfer from an AV spanner. For the axisymmetric disk (radius $R$) centered on an AV spanner, the total OAM transferred to the disk can be calculated as $\rho \int_0^R (\oint_L \boldsymbol{r} \times \boldsymbol{v}_\perp \cdot \mathrm{d}l) \cdot \mathrm{d}r$, where $L = 2\pi r$ is the circumference of a circle with the radius $r$. For the irrotational particle vibration velocity $\boldsymbol{v}_\perp$ in the AV field, $\oint_L \boldsymbol{v}_\perp \mathrm{d}l$ is equal to zero for an arbitrary radius $r$ based on Stokes' theorem, and then it can be transformed to $\oint_L \boldsymbol{v}_\perp \mathrm{d}l = r \int_0^{2\pi}(-v_i \sin\theta + v_j \cos\theta)\mathrm{d}\theta = 0$ for $\boldsymbol{v}_\perp = (v_i, v_j)$. Therefore, the OAM of a point on the circumference $L$ can be written as $\boldsymbol{r} \times \boldsymbol{v}_\perp = (r\cos\theta, r\sin\theta) \times (v_i, v_j) = v_i r \sin\theta + v_j r \cos\theta$ for $\boldsymbol{r}=(r\cos\theta, r\sin\theta)$. By taking curvilinear integral over $L$, we get $\oint_L (\boldsymbol{r} \times \boldsymbol{v}_\perp) \cdot \mathrm{d}l = r\int_0^{2\pi}(-v_i \sin\theta + v_j \cos\theta) \cdot \mathrm{d}\theta = 0$, indicating that the total OAM around the circumference $L$ is zero. Therefore, the total OAM transferred from the AV spanner to the disk is proved to zero. This suggests that the disk rotation in Volke-Sepúlveda's study [24] should not be considered as the result of the OAM transferred from the AV spanner. Contrary to the OAM, the acoustic radiation force with the direction of the wave vector $\boldsymbol{k}$ is not affected by the irrotationality of the instantaneous vibration velocity $\boldsymbol{v}$, which is more applicable for the analysis of object rotation for an AV spanner than the OAM. For example, in the object manipulation using an AV spanner, the size of the object in a human body (e.g. kidney stone in lithotripsy) is at millimeter-level, which is larger than the acoustic wavelength at the frequency of MHz. Therefore, the motion state of the object cannot be predicted precisely if only the OAM transfer from the AV spanner is considered. However, the analysis of the radiation torque exerted on the object as a whole shows its significance in the practical application.

In addition, in order to further confirm that the rotation of the coaxially placed disk is not the result of the OAM transfer, the OAM distributions of AVs with different topological charges were simulated and the corresponding total OAMs transferred to the disks of various radii were also calculated. The cross-sectional OAM distributions of AVs for $l$=1 to 4 at $z$=15 cm are illustrated in Fig. 6. For each $l$, a symmetric distribution is clearly displayed with an OAM zero at the vortex core. With the increase of the radius, obvious OAM enhancements occur with several distinct momentum peaks in opposite polarities. Meanwhile, with the increase of the topological charge, the OAM peaks move outward obviously to form an expanded AV spanner with a relative low-level OAM in the center region. Even though the OAMs at different positions might be quite different within an arbitrary radius $R$, the total OAM transferred to the disk ($\rho \int_0^R (\oint_L \boldsymbol{r} \times \boldsymbol{v}_\perp \cdot \mathrm{d}l) \cdot \mathrm{d}r$) can also be proved to be 0 for different topological charges, which further verifies that no OAM can be transferred from the AV spanner to the coaxially placed disk.

Moreover, in Jiang's study [32], the acoustic resonance in a planar layer of half-wavelength thickness was proposed to twist wave vectors of an incoming plane wave into a spiral phase dislocation of an outgoing vortex beam with the OAM. The principle for producing the spiral phase was deduced using the structure consisting of eight fan-like sections of resonators. The generation of an AV beam was demonstrated by numerical simulations and experimental measurements of cross-sectional distributions of pressure amplitude and phase. However, with the proposed structure of the assembled layer, only $l$=1 was realized and the existence of the OAM transfer from the AV was not certified theoretically or experimentally.

Although the radius dependences of radiation torque for AVs with various topological charges were qualitatively studied, some discrepancies are observed between theoretical and experimental results. One potential error is that, because the size of experimental sources was not much smaller than the wavelength, the simplified assumption of acoustic point source is not accurate completely. Further, the perfect generation of an AV beam cannot be realized because of the uncertain factors of the pressure consistency and the phase accuracy of sources. In addition, the experimental measurements are also influenced by the inhomogeneity of materials, such as the roughness of the sponge, the nonlinear elasticity of the rubber hose, the acoustic reflection of the bracket, the incomplete acoustic absorption below the source plane, and so on.

# V. CONCLUSION

Based on the analyses of the wave front and the wave vector for AVs generated by the phased-coded approach using a circular array of point sources, the topological-charge dependences of the radiation torque exerted on objects with different sizes were investigated theoretically and experimentally. It is shown that a more concentrated AV spanner with a smaller radius and a higher peak pressure can be generated by a little topological charge. The rotation of the axisymmetric disk centered on an AV spanner is driven by the exerted radiation torque without the contribution of the OAM transfer. The radiation torque is inversely associated with the topological charge for a small-radius object located in the center of an AV. For a larger object, a higher topological charge should be selected to enhance the rotation effect of the bigger AV. With the laser-displacement indicator, the proposed theory is verified by the experimental measurements of the topological-charge dependences of radiation torque for disks with different radii. It is demonstrated that, for an object covering the vortex center, the total OAM transfer might be 0, and the object rotation is the result of the exerted radiation torque of the AV spanner. The favorable results suggest that the radiation torque of an AV spanner is more applicable to understanding the rotation of objects, especially for the object with a feature size at the wavelength-scale (e.g. kidney stone in lithotripsy). Although only the low-frequency sound is employed in this study, the conclusions can be extended to the high-frequency ultrasound in water [26, 33], which might enable more potential applications by noninvasively manipulating objects inside body.

## ACKNOWLEDGMENTS

We would like to thank Professor Jianchun Cheng, Dr. Likun Zhang and Prof. Yafei Dai for their assistant in this research. This work was supported by the National Natural Science Foundation of China (Grant Nos. 11474166 and 11604156), the Natural Science Foundation of Jiangsu Province (No. BK20161013), the Postdoctoral Science Foundation of China (No. 2016M591874), the Postgraduate Research & Practice Innovation Program of Jiangsu

## Figure captions

FIG. 1. Schematic diagram of the AV generation and scanning measurement system.

FIG. 2. Axial profiles (line 1) and cross-sectional distributions of pressure (line 2) and phase (line 3) for AV spanners with the topological charges of 1, 2, 3 and 4 at $z$=15 cm.

FIG. 3. Normalized radial pressure distributions of AV spanners with different topological charges at $z$=15 cm.

FIG. 4. Radius dependences of the radiation torque and the deflection angle for AV spanners with different topological charges at $z$=15 cm.

FIG. 5. Sketch map of radiation torque measurement for the AV spanners with a laser-displacement (angle) indicator.

FIG. 6. Cross-sectional distributions of OAM for AV spanners with the topological charges of (a) 1, (b) 2, (c) 3 and (d) 4 at $z$=15 cm, which are normalized by the maximum value for $l$=4.

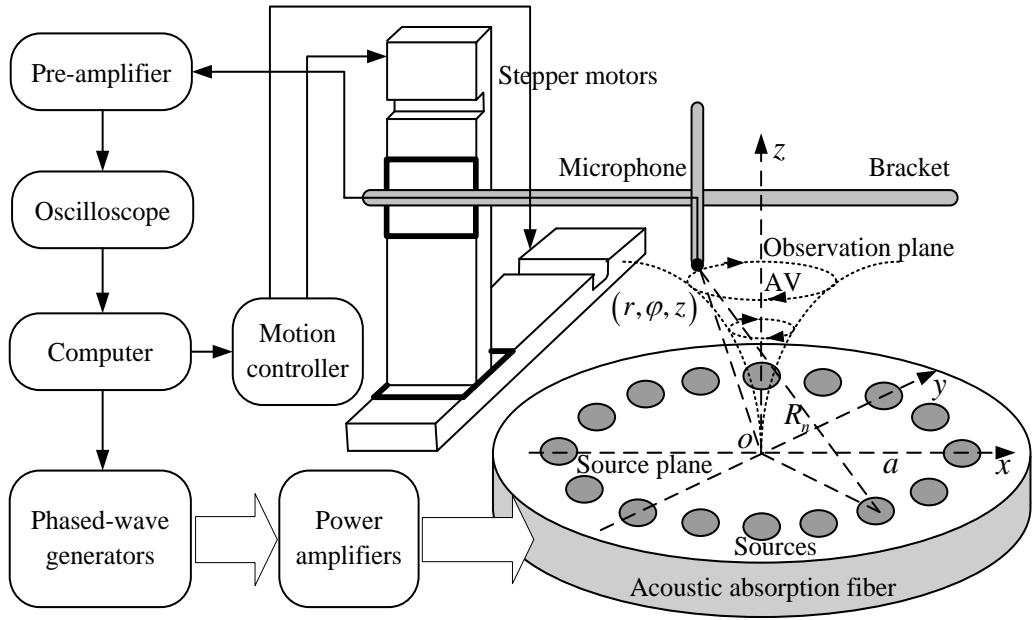

FIG. 1. Schematic diagram of the AV generation and scanning measurement system.

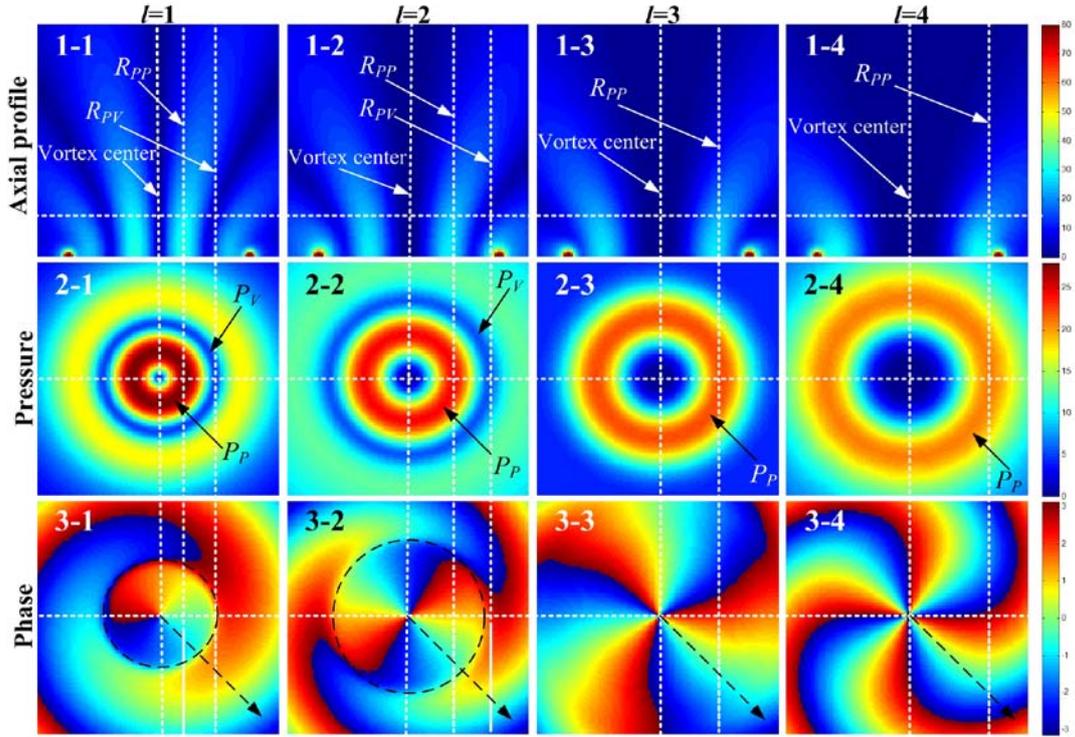

FIG. 2. Axial profiles (line 1) and cross-sectional distributions of pressure (line 2) and phase (line 3) for AV spanners with the topological charges of 1, 2, 3 and 4 at $z$=15 cm.

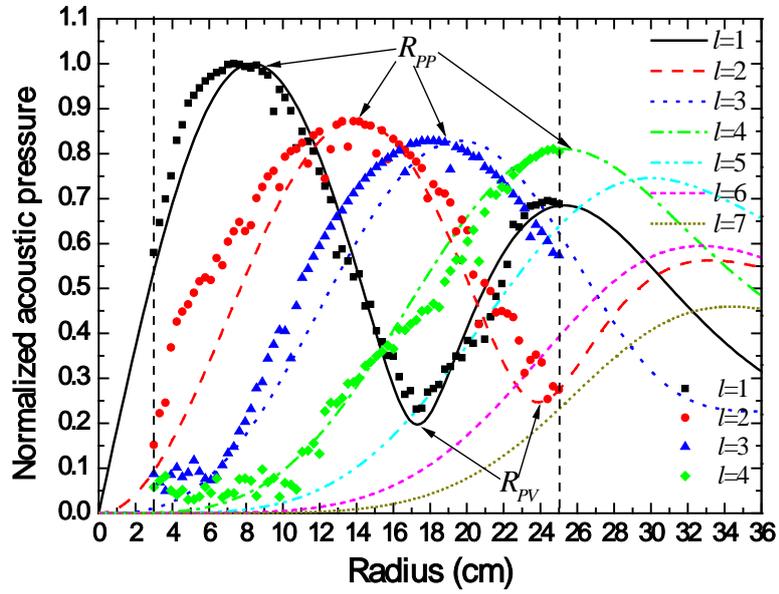

FIG. 3. Normalized radial pressure distributions of AV spanners with different topological charges at *z*=15 cm.

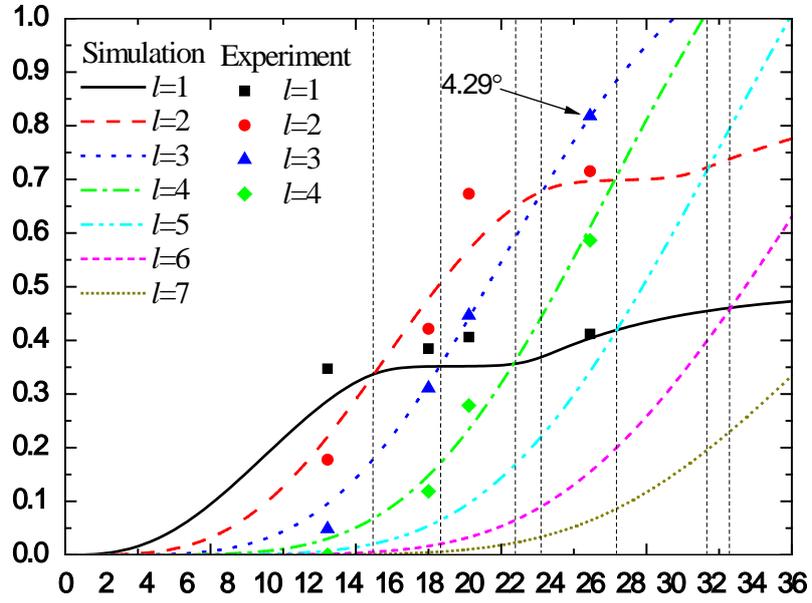

FIG. 4. Radius dependences of the radiation torque and the deflection angle for AV spanners with different topological charges at *z*=15 cm.

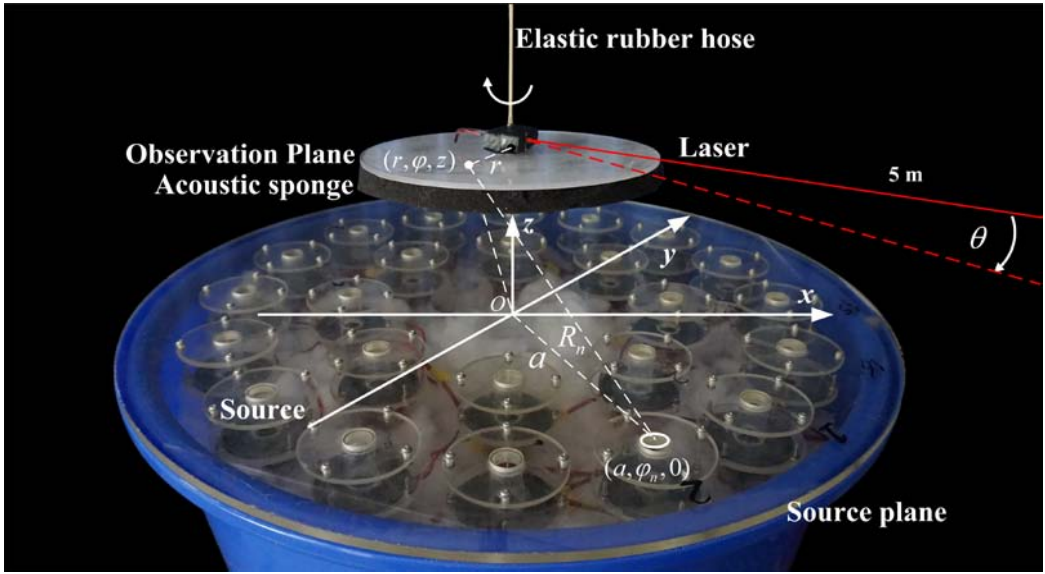

FIG. 5. Sketch map of radiation torque measurement for AV spanners with a laser-displacement (angle) indicator.

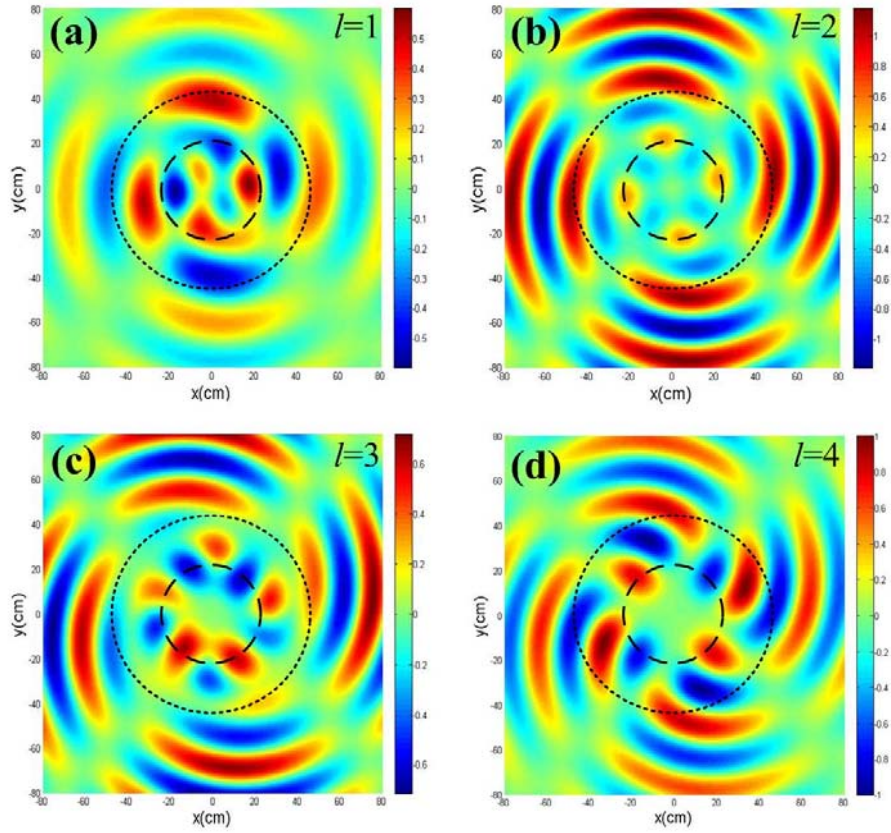

FIG. 6. Cross-sectional distributions of OAM for AV spanners with the topological charges of

(a) 1, (b) 2, (c) 3 and (d) 4 at *z*=15 cm, which are normalized by the maximum value for *l*=4.